\documentclass[seceq,jjapprn]{jjap}
\usepackage[dvips]{epsfig}

\title{Application of the Mesh Experiment for the Back-Illuminated
CCD: I. Experiment and the Charge Cloud Shape}

\author{Emi {\sc Miyata}, Masami {\sc Miki}, Junko {\sc
Hiraga}, Hirohiko {\sc Kouno}, Kazutoshi {\sc Yasui},
Hiroshi {\sc Tsunemi}, Kazuhisa {\sc Miyaguchi}$^{2}$, and Kouei
{\sc Yamamoto}$^{2}$ }

\inst{
Department of Earth \& Space Science, 
Graduate School of Science, Osaka University, \\
1-1 Machikaneyama, Toyonaka, Osaka 560-0043, Japan \\
$^1$ Solid State Division, Hamamatsu Photonics K.K., \\
1126-1 Ichino-cho, Hamamatsu City 435-8558, Japan
}

\abst{ We have employed a mesh experiment for back-illuminated (BI)
CCDs. BI CCDs possess the same structure to those of FI CCDs.  Since
X-ray photons enter from the back surface of the CCD, a primary charge
cloud is formed far from the electrodes. The primary charge cloud
expands through diffusion process until it reaches the potential well
that is just below the electrodes. Therefore, the diffusion time for the
charge cloud produced is longer than that in the FI CCD, resulting a
larger charge cloud shape expected.

The mesh experiment enables us to specify the X-ray point of interaction
with a subpixel resolution. We then have measured a charge cloud shape
produced in the BI CCD. We found that there are two components of the
charge cloud shape having different size: a narrow component and a broad
component.  The size of the narrow component is $2.8-5.7\,\mu$m in unit
of a standard deviation and strongly depends on the attenuation length
in Si of incident X-rays.  The shorter the attenuation length of X-rays
is, the larger the charge cloud becomes. This result is qualitatively
consistent with a diffusion model inside the CCD. On the other hand, the
size of the broad component is roughly constant of $\simeq 13\,\mu$m and
does not depend on X-ray energies. Judging from the design value of the
CCD and the fraction of each component, we conclude that the narrow
component is originated in the depletion region whereas the broad
component is in the field-free region.}

\kword{charge-coupled device, X-ray event, split event, subpixel spatial
resolution, mesh experiment}

\begin{document} 
\maketitle 
\sloppy

 \section{Introduction}

 A charge-coupled device (CCD) is widely used in an optical imaging as
 well as an X-ray imaging. When an X-ray photon is photoabsorbed in the
 CCD, it generates a primary charge cloud consisting of thousands of
 electrons. The size of the primary charge cloud is 73$\times(E/2.3
 \,{\rm keV})^{1.75}$\,nm where $E$ is energy of X-ray photon in unit of
 keV~\cite{charge_cloud}. The primary charge cloud expands through
 diffusion process until they reach the potential well of the CCD
 pixel. The final charge cloud after diffusion process is collected into
 several pixels forming various types of event pattern depending on how
 they split.

 We have developed a new technique ``mesh experiment'' which enables us
 to restrict the X-ray point of interaction with a subpixel
 resolution~\cite{tsunemi97}.  Hiraga {\it et al.}~\cite{hiraga98}
 investigated the event pattern produced and measured the charge cloud
 shape produced in the front-illuminated (FI) CCD for the first
 time. They found that a charge cloud shape could be well represented by
 a Gaussian function.  They also obtained the standard deviation,
 $\sigma$, of the final charge cloud to be $0.7 \sim 1.5 \,\mu$m for
 $1.5\sim 4.5$\,keV X-rays by using a FI CCD.  Based on this experiment,
 they confirmed that there are three parameters tightly coupled
 together~\cite{tsunemi00}; an X-ray point of interaction within a
 pixel, an X-ray event pattern and a charge cloud shape.  Any two
 parameters can determine the third one.  The event pattern is quite
 easily noticed while the charge cloud shape is difficult to be
 measured.  Currently, they can separately measure it only by using the
 mesh experiment.  Therefore, they can determine the X-ray point of
 interaction with much better spatial resolution than the pixel size.
 They obtained the position resolution of 0.7\,$\mu$m using the pixel
 size of 12\,$\mu$m if an X-ray photon became a split pixel
 event. However, due to its relatively small size of the charge cloud,
 the fraction of split pixel events whose point of interaction can be
 improved is less than 10\,\% of the total events.

 In this paper, we applied the mesh experiment to the back-illuminated
 (BI) CCD.  Figure~\ref{fig:fi_cross_section} shows the cross-section
 and the potential profile of FI CCDs. In the case of FI CCDs, since
 X-ray photons enter from the front side where there are electrodes.  A
 major part of X-rays are photo-absorbed close to electrodes, resulting
 a final charge cloud to be relatively
 small. Figure~\ref{fig:bi_cross_section} shows the cross-section and
 the potential profile of BI CCDs.  In the case of BI CCDs, a substrate
 of CCD, shown in Fig~\ref{fig:fi_cross_section} (a), is etched from the
 back toward the depletion layer (thinning process) in order to achieve
 high quantum efficiency for blue lights and low energy X-rays. After
 the thinning process, almost all of the substrate has been removed.
 The thickness of the BI CCD employed is $\simeq20\,\mu$m.  Since X-ray
 photons enter from the back surface of the CCD shown in
 Fig~\ref{fig:bi_cross_section}, a primary charge cloud is formed far
 from the electrodes.  The primary charge cloud expands through
 diffusion process until they reach the potential well which is just
 below the electrodes. Therefore, the diffusion time for the charge
 cloud generated by X-rays is longer than those in the FI CCD, resulting
 that a larger charge cloud shape is expected.

 Low energy X-rays, however, are photo-absorbed in the field-free region
 of BI CCDs. The primary charge cloud expands due to a diffusion process
 toward the electrodes as well as the back surface. The charge traveled
 to the back surface will be lost due to a recombination process. In the
 production of BI CCDs, accumulation is the most significant process
 that gives an internal potential to the back surface of CCDs and repels
 the signal charge to the electrodes~\cite{accumulation1}.  In the case
 of the BI CCD employed, the accumulation process is performed by ion
 implantation~\cite{accumulation2}.  As shown in
 Fig~\ref{fig:bi_cross_section} (b), the potential at close to the back
 surface is locally lower than that at the inner region caused by the
 accumulation process which enables us to collect the charge generated
 close the back surface.

 \section{Experiment and Results}

 The basic idea and the algorithm of data reconstruction for the mesh
 experiment are given in Tsunemi {\it et al.}\cite{tsunemi97}.  We
 performed mesh experiment employing a BI CCD (S7170) fabricated by
 Hamamatsu Photonics K.K. The BI CCD has a two-phase electrode and
 full-frame transfer type. It possesses 512$\times$512 active pixels
 with a size of 24\,$\mu$m square. The mesh made of gold has small holes
 of $2.1\,\mu$m diameter with spaced 48\,$\mu$m apart. The pitch of mesh
 holes is twice larger than that of the CCD pixel.  The mesh is placed
 just $5\,$mm above the CCD.  We employed the 21\,m long X-ray beam line
 in our laboratory. We used the X-ray generator of Ultra-X18, fabricated
 by RIGAKU, with Al, Mo, and Ti targets in order to irradiate the
 characteristic emission line of each target: Al-K\,(1.5\,keV),
 Mo-L\,(2.3\,keV), and Ti-K\,(4.5\,keV).  We applied the voltage of 6,
 5, and 7\,kV for Al, Mo, and Ti target, respectively.  We used a
 mechanical shutter made of stainless steel having thickness of
 60\,$\mu$m to control the X-ray beam so that the pile-up would not
 become a serious effect.  We drove the BI CCD with our newly-developed
 system named {\sl E-NA} system~\cite{e-na}. The CCD analog data were
 processed by an integration-type circuit~\cite{ssc_em} and digital data
 were acquired with a VME system. We controlled the operating
 temperature of the CCD to be $-100\,^\circ$C during the experiment.

 Figure~\ref{fig:spectra} shows X-ray spectra of single-pixel events
 obtained with three targets employed. The readout noise of our system
 is $\simeq$\,20\,e$^-$rms.  Each spectrum shows the characteristic
 emission line superimposed on the continuum emission. We should note
 that the peak channel of Mo-L emission line is close to that of Al-K
 emission line whereas the energy is quite different.

 We selected events between two dashed lines shown in
 Fig~\ref{fig:spectra} in the subsequent analysis.

 \subsection{Event pattern}

 Figure~\ref{fig:single_raw_image} shows a part of a raw image obtained
 with single-pixel events. We can see a clear moir\'{e} pattern. Based
 on the moir\'{e} pattern obtained, we can determine the geometrical
 relation between the mesh and the CCD.  Details of the method are given
 in Tsunemi {\it et al.}~\cite{tsunemi98}.

 Once we find their relations, we can determine the X-ray point of
 interaction within a pixel for all X-ray events. We thus rearranged
 each X-ray event with a subpixel resolution and investigated the
 distribution of single-pixel event, two-pixel split event, three- and
 four-pixel split event and more extended events as a function of the
 point of interaction inside a pixel. Figure~\ref{fig:rp} shows their
 X-ray intensity maps in 3$\times$3 representative pixels (RP) for Mo-L
 X-rays. Each black dot corresponds to an individual X-ray event. When
 X-ray photons enter the center of a pixel, a major part of them produce
 single-pixel events shown in (a). When the X-ray point of interaction
 becomes close to the horizontal or vertical boundary of each pixel, the
 horizontally or vertically two-pixel split events are formed shown in
 (b) or (c). When an X-ray photon enters the corner of each pixel, it
 produces three- or four-pixel split event shown in (d). This tendency
 is consistent with that obtained for the FI
 CCD~\cite{tsunemi99}. However, we found that the regions forming the
 three- and four-pixel split events become larger and extend into inner
 region than FI results (e.g. Figure~7 of Tsunemi {\it et
 al.}~\cite{tsunemi98}) whereas the pixel size of the BI CCD employed is
 24\,$\mu$m that is two times larger than that of the FI CCD. This
 suggests that a charge cloud size produced in the BI CCD is much larger
 than that of the FI CCD. This fact is clarified in Fig~\ref{fig:rp} (e)
 since a charge cloud spreads into more than four pixels even if an
 X-ray enters the center of a pixel.

 \subsection{Absorption structure of BI CCDs}

 As shown in Fig~\ref{fig:fi_cross_section}, X-ray photons enter from
 the electrodes in the case of FI CCDs. The thickness of electrodes is
 not uniform and some part of electrodes is overlapped, resulting the
 detection efficiency within a pixel to be far from
 uniform~\cite{tsunemi97, kyoshita, pivovaroff98, mosccd}.  It makes it
 difficult to make the response matrix of FI CCDs.

 As to the BI CCDs, on the other hand, X-ray photons enter from the back
 surface where the thickness of the field-free region is much thinner
 than those of FI CCDs shown in Fig~\ref{fig:bi_cross_section}. The
 potential created by the accumulation process also drives the charge
 generated in the field-free region toward the electrodes, resulting
 higher detection efficiency for low energy X-rays. Since the thickness
 of the back surface of CCDs is uniform, we can expect a uniform
 detection efficiencies within pixels for BI CCDs.
 Figure~\ref{fig:all_grades} shows the X-ray intensity map extracted
 from all X-ray events in 3$\times$3 RPs for Mo-L X-rays. There is no
 significant variation in each image, demonstrating the uniform
 detection efficiency of the BI CCD. We also confirmed uniformity for
 Al-K and Ti-K X-rays.  The non-uniformities in detection efficiency are
 $\simeq$ 2.5, 2.0, and 7.2\,\% in a standard deviation for Al-K, Mo-L,
 and Ti-K X-rays, respectively. This uniformity is a large advantageous
 point of BI CCDs over FI CCDs.  We should note that the uniformity of
 Ti-K X-rays is lower than those of other X-rays. It must be due to the
 fact that the mean attenuation length of Ti-K X-rays is long enough for
 X-ray point of interaction to be affected by electrodes.

 \subsection{Amount of charge collected in a center pixel}

 When X-rays enter the CCD at the position of $(X_p, Y_p)$, the output
 of the $n_{\rm th}$ pixel of the CCD, $P_n (X_p, Y_p)$, is written as
 \[
  P_n (X_p, Y_p) = \int_{X_n}^{X_{n+1}} dX \int_{Y_n}^{Y_{n+1}} dY\
  f(X-X_p, Y-Y_p)
 \]
 where ($X_n, X_{n+1}, Y_n, Y_{n+1}$) denotes the boundary of the $n_{\rm
 th}$ pixel and $f(X, Y)$ is the charge cloud shape just before the
 collection by the potential well. The actual output of $n_{\rm th}$
 pixel, $D_n (X_{in}, Y_{in})$, is a convolution between $P_n (X_p,
 Y_p)$ and the mesh hole,  which can be described by

 \begin{eqnarray}
  \label{eq_dn}
  D_n (X_{in}, Y_{in}) &=& \int dX_p \int dY_p\ P_n (X_p, Y_p)
   \ H(X_{in}-X_p, Y_{in}-Y_p) \\
   & = & \int_{X_n}^{X_{n+1}} dX \int_{Y_n}^{Y_{n+1}} dY\
    f \otimes H (X-X_{in}, Y-Y_{in})
 \end{eqnarray}

 \noindent where ($X_{in}, Y_{in}$) is the center of mesh hole, $H(X,
 Y)$ is a typical hole shape of the mesh, and $f \otimes H$ represents
 the convolution between $f$ and $H$.

 We can experimentally measure $D_n (X_{in}, Y_{in})$ by using the RP
 generated with all X-ray events. Figure~\ref{fig:dn} show the surface
 profiles and projected profiles of $D_n$ in 3$\times$3 pixel regions
 where the center pixel is $n_{\rm th} $ pixel for Mo-L, Al-K, and Ti-K
 X-rays.

 Hiraga {\it et al.}~\cite{hiraga98} measured $D_n$ for the FI CCD for
 the first time and found a flat-top feature around the center of
 $n_{\rm th}$ pixel. This is due to the fact that the charge collected
 in the $n_{\rm th}$ pixel is almost constant, forming single-pixel
 events, when a mesh hole is well within the pixel.  $D_n$ obtained with
 Ti-K for the BI CCD has relatively flat top at the center of pixel. On
 the other hand, $D_n$ shows a sharp structure even near the center of
 pixel in Mo-L and Al-K X-rays, suggesting that a charge splits into
 neighboring pixels even if X-rays enter near the center of the pixel.
 We can again expect that a charge cloud size produced in the BI CCD is
 much larger than that of the FI CCD. It is interesting to note that a
 flat-top feature obtained only with Ti-K X-rays suggests that a size of
 the charge cloud of Ti-K X-rays is smaller than those obtained with
 Mo-L and Al-K, which is in contrast to those obtained with FI
 CCDs~\cite{tsunemi99}.

 \subsection{Charge cloud shape}

 Hiraga {\it et al.}~\cite{hiraga98} calculated a charge cloud shape by
 differentiating $D_n$ (we hereafter refereed to the differential
 method). As shown in Fig~\ref{fig:dn}, however, we cannot obtain a
 charge cloud shape with the differential method since $D_n$ does not
 show a flat-top feature for the BI CCD. We then employed a different
 approach to obtain a charge cloud shape.

 As written in eq~\ref{eq_dn}, $D_n$ is an integration of convolution
 between a charge cloud shape and a mesh hole within a CCD pixel. We
 directly calculate their convolution to reproduce $D_n$ for each energy
 assuming a charge cloud shape to be a two-dimensional axial-symmetric
 Gaussian function~\cite{hiraga98} (the integral method, hereafter).  We
 should note that we have applied the integral method to the FI data and
 obtained consistent results to those obtained by the differential
 method. We found, however, a difficulty to reproduce $D_n$ with the
 integral method. The problem is that our method cannot reproduce a
 sharp component and a tail component, simultaneously. Such a tail
 component is detected for all energies whereas it was not detected with
 FI CCD~\cite{hiraga98}. The presence of an tail component suggests a
 large size of a charge cloud which must be much larger than that
 forming a sharp component. As shown in Fig~\ref{fig:rp} (e), the events
 spreading more than 4 pixel is uniformly distributed within a
 pixel. These events should account for such tail components.

 We thus introduce an extra component to fit the data. We employed two
 components in the charge cloud shape having different size in both
 axes: a narrow component for the sharp structure and a broad component
 for the tail structure. In this way, we found that the two-component
 model in the integral method well reproduced $D_n$ obtained.  The model
 functions of $D_n$ are shown in Fig~\ref{fig:dn_model}.
 Table~\ref{table:size} summarizes the charge cloud sizes of two
 components for all energies.  Figure~\ref{fig:charge_cloud_size} shows
 the size of the charge cloud for both components as a function of the
 X-ray attenuation length in Si.

 The broad component of the charge cloud is $2.5-5$ times larger than
 that of the narrow component.  The size of the narrow component
 decreases as the attenuation length increases whereas that of the broad
 component is relatively constant of $\sim 13\,\mu$m.  Comparing the FI
 results~\cite{tsunemi99}, the sizes for the narrow and the broad
 component are 2$-$8 times larger and an order of magnitude larger,
 respectively.  The number of X-ray events of the broad component
 is $\sim$\,3 times larger than that of the narrow component.

 \section{Discussion}

 \subsection{Non-linearity of pulse height versus energy}

 As shown in Fig~\ref{fig:spectra}, the pulse height of each
 characteristic X-ray line does not linearly increase with energy.
 Figure~\ref{fig:nonlinearity} shows their relationship between the
 pulse height and the incident X-ray energy.  The pulse height of Mo-L
 is $\simeq 30$\,\% lower than that expected from the linear
 relationship based on Al-K and Ti-K X-rays as shown by a solid line.
 This is quite a different characteristics of the BI CCD to the FI CCD.

 The attenuation length of Mo-L X-rays is only 2.2\,$\mu$m that is the
 shortest among X-rays employed. Therefore, most of Mo-L X-rays are
 absorbed close to the accumulation region, where the density of
 impurities is an order of magnitude larger than that in the depletion
 region, leading to the recombination between impurities and signal
 charge. We thus suppose that the non-linearity of pulse height
 versus energy is caused by an incomplete collection of the signal
 charge produced in the accumulation or some part of the field-free
 region.

 \subsection{Origin of the two components of charge cloud}\label{two_comp}

 We found that there were two components in a charge cloud shape
 generated by the BI CCD.  A narrow components has a $\sigma$ of $2.8 -
 5.7\, \mu$m which strongly depends on the incident X-ray energy.  The
 size of a broad component is relatively constant of $\simeq
 13\,\mu$m having less dependence on the X-ray energy whereas such
 a component is not appeared in the FI CCD~\cite{hiraga98}.

 As to BI CCDs, X-rays entering from the back surface of the CCD form a
 primary charge cloud far from the electrodes. The primary charge cloud
 expands through diffusion process until they reach the potential well
 which is just below the electrodes. Therefore, a diffusion time for the
 charge cloud produced by low energy X-rays is longer than that by high
 energy X-rays, resulting a larger size of charge cloud shape for low
 energy X-rays. This relation is consistent with that obtained for a
 narrow component.  However, we found that the size of the broad
 component is almost independent of the incident X-ray energy.

 As shown in Figs~\ref{fig:fi_cross_section} (b) and
 \ref{fig:bi_cross_section} (b), the potential profiles of FI and BI
 CCDs are different in the field-free region. A charge generated in the
 field-free region can be collected only through the diffusion process
 in FI CCDs. On the other hand, the potential created by the
 accumulation drives the charge generated in the field-free region
 toward electrodes in the BI CCD.  Therefore, a charge generated both in
 the depletion region and in the field-free region can be collected in
 the BI CCD. With taking into account the difference in the charge cloud
 size for both components, we suppose that an X-ray photo-absorbed in
 the depletion region can be detected as a narrow component whereas
 an X-ray photo-absorbed in the field-free region can be detected as
 a broad component.

 Miyata {\it et al.}~\cite{ssc_em} measured the thickness of the
 depletion depth of FI CCD having similar wafer as that of the BI CCD
 employed to be $\simeq 4\,\mu$m. Since the thickness of the BI CCD
 employed is 20\,$\mu$m, thickness of the field-free region is $\simeq
 16\,\mu$m.  This value is in good agreement with the charge cloud size
 of the broad component if the charge cloud expands spherically in the
 field-free region. Moreover, as shown in Table~\ref{table:size}, the
 fraction of the broad component obtained is $\sim$\,3 times larger than
 that of the narrow component, which is also consistent with the ratio
 of thickness between the field-free region and the depletion
 region. These facts strongly support that a broad component is 
 originated in the field-free region.

 \subsection{Asymmetry of the charge cloud shape}

 The charge cloud size obtained with the BI CCD is shown in
 Table~\ref{table:size}. Both for the narrow and broad components, the
 charge cloud shapes are almost point symmetric. The difference in both
 directions is less than 5\,\%.  This is also noticed that  $D_n$ obtained
 for three energies shown in Fig~\ref{fig:dn} show a quite
 point-symmetric shape.

 Hiraga {\it et al.}~\cite{hiraga98} pointed out that a charge cloud
 shape obtained for the FI CCD shows an asymmetric shape. The size in
 unit of the standard deviation for the x-direction was $1.4\,\mu$m
 whereas that for the y-direction was $0.7-0.8\,\mu$m. Tsunemi {\it et
 al.}~\cite{asymmetry} further investigated the possible explanations
 about the asymmetry of the charge cloud shape. They conclude that the
 asymmetry is due not to the experimental setup, such as the asymmetric
 mesh hole, but to the asymmetry of the electric field inside the
 CCD. Similar asymmetry has obtained for some other FI
 CCDs~\cite{tsunemi99, kyoshita}. Therefore, the asymmetric shape of the
 charge cloud is commonly found in FI CCDs.

 On the other hand, we obtained a symmetric shape of the charge cloud
 generated inside the BI CCD for all three X-rays. The back surface of
 CCDs are uniform and is normally applied to be zero voltage. At the
 front side, there are four electrodes per pixel in our case and they
 are slightly overlapped to each other as shown in
 Fig~\ref{fig:fi_cross_section}\,(a). Moreover, at the front side, there
 is another gate structure, ``channel stop'', which forms the pixel
 boundary along the y-direction. All these gate structures generate the
 electric field not to be uniform near the front electrodes as pointed
 out by Tsunemi {\it et al.}\cite{asymmetry}.  The symmetry of the
 charge cloud obtained with BI CCD suggests that a charge cloud must not
 be affected by the electrodes significantly because it is generated far
 from the electrodes.

 \section{Conclusion}

 We have performed the mesh experiment with the BI CCD and measured the
 charge cloud size for the first time. We found that there are two
 components of the charge cloud shape to reproduce the data obtained all
 for Mo-L, Al-K and Ti X-rays: a narrow component and a broad component.
 The narrow component possesses a $\sigma$ of $2.8 - 5.7\, \mu$m which
 strongly depends on the attenuation lengths of incident X-ray
 energies. The broad component, on the other hand, possesses much larger
 $\sigma$ than that of the narrow component. It is $\simeq 13\,\mu$m and
 almost independent of the attenuation length of X-rays. The accumulation
 technique to fabricate BI CCDs is not employed in the FI CCDs
 and the charge produced even in the field-free region can be collected
 toward electrodes. We thus suppose that the narrow and the broad
 components are originated in the depletion region and the field-free
 region, respectively. This hypothesis is consistent with both by the
 design value of the CCD and by the fraction of X-ray events for both
 components.

 We have found that the linearity between the incident X-ray energies
 and the pulse height is not good. The pulse height of Mo-L X-rays is
 $\simeq 30$\,\% lower than that expected.  This might be explained by the
 fact that the effect of the recombination with the impurities is
 significant in the accumulation region since the attenuation length of
 Mo-L is shortest among those X-rays employed.

 \acknowledgement

 J.H. is partially supported by JSPS Research Fellowship for Young
 Scientists, Japan. This work is partly supported by the Grant-in-Aid
 for Scientific Research by the Ministry of Education, Culture, Sports,
 Science and Technology of Japan (13874032, 13440062).

 \begin{halftable}
  \caption{Charge cloud shape obtained with the BI CCD for Mo-L, Al-K, and
  Ti-K X-rays.}
  \label{table:size}
  \begin{halftabular}{@{\hspace{\tabcolsep}%
   \extracolsep{\fill}}lccc} \hline
    & Mo-L & Al-K & Ti-K\\ \hline
   Energy [keV] & 2.3 & 1.5 & 4.5 \\
   Attenuation length [$\mu$m] & 2.2 & 7.9 & 13.5 \\\hline
   \multicolumn{4}{l}{Narrow component}\\\hline
   \ \ $\sigma_{nx}$ & $5.3 \pm 0.3$ & $4.3 \pm 0.2$ & $2.91 \pm 0.08$ \\
   \ \ $\sigma_{ny}$ & $5.8 \pm 0.3$ & $4.4 \pm 0.2$ & $2.99 \pm 0.09$ \\
   \ \ $S_{nx}$ & $5.3 \pm 0.3$ & $4.2 \pm 0.1$ & $2.83 \pm 0.08$ \\
   \ \ $S_{ny}$ & $5.7 \pm 0.3$ & $4.3 \pm 0.2$ & $2.91 \pm 0.09$ \\ \hline
   \multicolumn{4}{l}{Broad component}\\\hline
   \ \ $\sigma_{wx}$ & $13.4 \pm 0.3$ & $13.2 \pm 0.3$ & $11.9 \pm 0.1$ \\
   \ \ $\sigma_{wy}$ & $13.9 \pm 0.3$ & $13.8 \pm 0.3$ & $12.5 \pm 0.1$ \\
   \ \ $S_{wx}$ & $13.4 \pm 0.3$ & $13.2 \pm 0.3$ & $11.9 \pm 0.1$ \\
   \ \ $S_{wy}$ & $13.9 \pm 0.3$ & $13.8 \pm 0.3$ & $12.5 \pm 0.1$ \\ \hline
   \multicolumn{4}{l}{Fraction of narrow component}\\\hline
   & $0.25 \pm 0.02$ & $0.32 \pm 0.01$ & $0.32 \pm 0.01 $ \\
   \hline
  \end{halftabular}
 \end{halftable}
 
 \begin{halffigure}
  \caption{(a) Cross-section and (b) potential profile of a FI CCD.}
  \label{fig:fi_cross_section}
 \end{halffigure}
 
 \begin{halffigure}
  \caption{Same as Fig~\ref{fig:fi_cross_section} but for a BI CCD.}
  \label{fig:bi_cross_section}
 \end{halffigure}

 \begin{halffigure}
  \caption{X-ray spectra obtained with targets of (a) Al, (b) Mo, and
  (c) Ti. Each characteristic emission lines are identified. The dotted
  lines show the ranges employed for analysis.}
  \label{fig:spectra}
 \end{halffigure}

 \begin{halffigure}
  \caption{A part of the raw image ($200 \times 200$ pixels) extracted
  from the single-pixel events of Mo-L X-rays. Each dot shows an X-ray
  event. A clear Moire pattern can be seen. Since the pitch of the mesh
  hole is two times that of the CCD pixel, we find X-ray events appeared
  in every 2$\times$2 pixels.}
  \label{fig:single_raw_image}
 \end{halffigure}

 \begin{halffigure}
  \caption{X-ray intensity map of (a) single-pixel events, (b)
  horizontally two-pixel split events, (c) vertically two-pixel split
  events in 3$\times$3 RPs, (d) three- and four-pixel split events, and
  (e) more extended events with the subpixel resolution. The dotted lines
  show the pixel boundary.}
  \label{fig:rp}
 \end{halffigure}

 \begin{halffigure}
  \caption{X-ray intensity map extracted from all X-ray events in
  3$\times$3 RPs for Mo-L.  The uniform detection efficiency can be
  found for BI CCD. Projected profiles onto X- and Y-axes are also
  shown.  The dotted lines show the pixel boundary.}
  \label{fig:all_grades}
 \end{halffigure}

 \begin{halffigure}
  \caption{Amount of charge collected in the $n_{\rm th}$ pixel, $D_n$,
  is shown in 3$\times$3 pixels for (a) Mo-L, (b) Al-K, and (c) Ti-K
  X-rays. Upper panel of each figure shows the 3-d picture of $D_n$ and
  projected profiles are shown in lower panel.}  \label{fig:dn}
 \end{halffigure}

 \begin{halffigure}
  \caption{Model of $D_n$ for  (a) Mo-L, (b) Al-K, and (c) Ti-K
  X-rays.}  \label{fig:dn_model}
 \end{halffigure}

 \begin{halffigure}
  \caption{Charge cloud size as a function of the X-ray attenuation
  length in Si for (a) the narrow component and (b) the broad component. }
  \label{fig:charge_cloud_size}
 \end{halffigure}
 
 \begin{halffigure}
  \caption{Relationship between the pulse height and incident X-ray
  energy. The solid line shows the linear relationship between them to
  reproduce Ti-K X-rays.}
  \label{fig:nonlinearity}
 \end{halffigure}

 \makefigurecaptions
 
\end{document}